\documentclass{ptapap}
\usepackage{amsmath}
\usepackage{amssymb}

\author{Bo\. zena Czerny}[CFT]
\author{Marzena \' Sniegowska}[TelAviv,CAMK]
\author{Agnieszka Janiuk}[CFT]
\author{Bei You}[Wuhan]
\affil[CFT]{Center for Theoretical Physics, Polish Academy of Sciences, Al. Lotnik\' ow 32/46, 02--668 Warsaw, Poland}
\affil[TelAviv]{School of Physics and Astronomy, Tel Aviv University, Tel Aviv 69978, Israel}
\affil[CAMK]{Nicolaus Copernicus Astronomical Center, Polish Academy of Sciences, ul. Bartycka 18, 00-716 Warsaw, Poland}
\affil[Wuhan]{Department of Astronomy, School of Physics and Technology, Wuhan University, Wuhan 430072, China}

\title{Accretion processes onto 
black holes: theoretical problems, observational constraints}
\headtitle{Accretion processes onto black holes}

\begin{document}

\maketitle

\begin{abstract}

We shortly summarize the standard current knowledge on the structure of the accretion flow onto black holes in galactic binary systems and in active galactic nuclei. We stress the similarities and differences between the two types of systems, and we highlight the complementarity of the data caused by these differences. We highlight some new developments and list the unsolved problems.   

\end{abstract}

\section{Introduction}

 Black holes are the key ingredients of the two very different types of objects. Some of the Galactic binary stars contain a black hole of the mass of the order of 10 solar mass as one of the components. These objects are sources of vigorous X-ray emission caused by the flow of the matter from the companion star towards the black hole \citep[see e.g.][]{McClintockBHB2006}. Much more massive, supermassive black holes ($10^5 - 10^{10}$ solar mass) are at the nuclei of all regular galaxies \citep[see e.g.][]{kormendy2013}. When there is enough interstellar material flowing towards the black hole, the nucleus becomes bright, with a strong non-stellar emission component, and it is classified as an active galactic nucleus (AGN) \citep[see e.g.][]{krolik1999}.

 The number of known galactic black holes (GBH) is relatively small, less than 200 \citep[see e.g.][]{ziolkowski2010,fortin2023} since we find them only in the Milky Way and in nearby galaxies like Large Magellanic Cloud. However, due to the relatively small distance, the quality of the observational data, particularly in the X-ray band is relatively high. Known AGN are much more numerous - one of the most recent quasar catalogs \citep{wu2022} lists over 700 000 objects. Quasars are just the most extreme versions of AGN. Some AGN are located in relatively close galaxies, like M87, where the black hole shadow was directly imaged \citep{M872019}. Our Galaxy - Milky Way - also contains a supermassive black hole, and this one was also imaged recently \citep{SgrA*2022}. However, it is only very weakly active, and we see this activity only due to a relatively small distance to the source.

\section{Key elements of accreting black holes and unwanted complications}

If we want to concentrate on the understanding of the accretion flow close to the black hole we must avoid sources that hide their centers from the observer. In some AGN the central parts are shielded by the dusty torus, in GBH by the outer rim of the disk. Some sources are also heavily obscured by the material flowing out from the central regions. Finally, some objects (particularly some AGN known as blazars) have strong spectacular jets which, when pointing towards us bind the observed since jet emission is relativistically boosted. These outflows are of course tightly related to the accretion process but they complicate considerably the analysis. In this review, we mostly concentrate on sources with an unshielded view of the black hole vicinity. 

The key ingredients of the system are thus a black hole and inflowing material. A black hole is simply characterized by the mass and spin (we usually assume the electric charge is zero, but see \citealt{zajacek2018}). Inflowing material usually has some angular momentum so it forms an accretion disk around a black hole. However, the flow can be dense and slow, forming a geometrically thin, optically thick Keplerian accretion disk \citep{SS1973}. The flow can be also fast, with material having no time to cool, and then it forms an optically thin, geometrically thick disk with a non-Keplerian angular momentum profile and considerable advection of energy down the horizon \citep{adaf1994}.

The inflow is usually accompanied by some outflow, in particular, it can have a form of a well-collimated jet. Two basic mechanisms of the jet formation are under consideration, as proposed by \citet{BZ1977}, and by \citet{BP1982}. We do not concentrate much on the sources with appearances strongly dominated by relativistic jets. We also concentrate mostly on highly active sources, as measured by their Eddington ratio, above 0.001.

\section{Broad band spectra of AGN and GBH}

\begin{figure}
  \centering
  \begin{minipage}{0.48\textwidth}
\includegraphics[width=\textwidth]{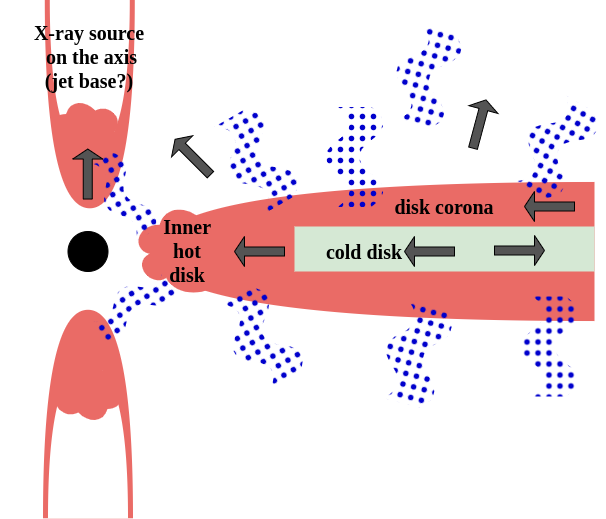}    
    \caption{Schematic universal picture of accreting black hole.}
    \label{fig:schem2}
  \end{minipage}
  \quad
  \begin{minipage}{0.45\textwidth}
    \includegraphics[width=\textwidth]{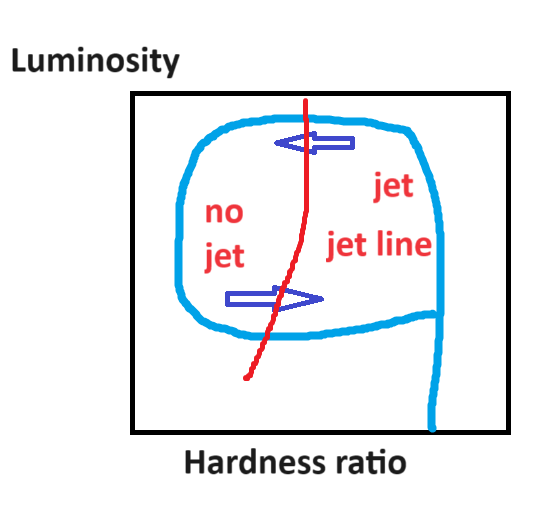}
    \caption{Schematic illustration of the evolution of X-ray nova during the whole cycle.}
    \label{fig:turtle}
  \end{minipage}
\end{figure}

The analysis of the broad spectra of GBH and AGN for relatively bright sources indicate the essential role of the following components: (i) cold Keplerian disk (ii) dusty torus or outer disk ring, in AGN and GBH, respectively, (iii) compact hot corona or extended hot inner flow (for high and very low Eddington ratio, respectively (iv) soft X-ray excess (v) jet. Apart from very low Eddington rate sources (below $\sim $ 0.01), the cold disk dominates energetically. Due to the systematic difference in the black hole mass, the characteristic temperature of the disk is lower in AGN (about $10^5$ K), and higher in GBH (about $10^7$ K). Therefore, this component is well seen in the optical/UV band for AGN \citep[e.g.][]{Capellupo2015,czerny2011}, and in X-ray band in the GBH \citep[e.g.][]{rodi2021}. Black hole spins determined from the disk spectral properties are frequently moderate \citep[see e.g.][]{zhao_MAXI_2021} but higher values are also found (see \citealt{McClintock2006}, \citealt{gou2009} or \citealt{zhao2021}). In hard X-ray we see a power law component related to the hot corona, and the soft X-ray excess is usually present in bright sources. The hot corona irradiates the cold disk and leads to the formation of relativistically smeared reflection component if the cold disk is present close to the black hole. X-ray properties related to the hot flow are very similar in AGN and GBH since the difference in the black hole mass does not affect these components.

\section{General picture of the inner flow}

The universal geometrical setup is illustrated in Fig.~\ref{fig:schem2}. The hot compact corona, with the electron temperature $\sim 10 ^9$ K, is most likely related to the inner jet, where efficient dissipation takes place, and the rest of the jet is not highly relativistic and powerful. There is no such dissipation in radio-loud jetted sources. Inner hot flow is extended in the case of very low Eddington sources, in extreme cases like Sgr A* at the center of our Galaxy there is no outer cold disk, while in bright sources cold disk actually approaches the innermost stable circular orbit (ISCO). Its location depends on the black hole spin. Warm corona forms where the coronal flow is dense and overlaps with the inner disk. Such a warm corona has a moderate temperature of $10^6 - 10^8$ K and, a relatively large optical depth of order of 10, and such a medium well explains the soft X-ray excess. This schematic picture did not include the disk's outer part where we can have a dusty/molecular torus (in AGN) and outer puffed disk ring (in GBH) which can shield the view of the inner disk for the highly inclined observers.

\section{Formation and disappearance of the jet}

In AGN we usually divide the sources into radio-quiet (non-jetted) and radio-loud (jetted) sources. However, it is important to notice that in GBH which undergoes X-ray novae outburst a given source in a matter of a few months strongly changes the accretion rate, from low to high and again back to a low rate, performing a complex evolution well illustrated in the diagram plotting the luminosity of the source against the hardness ratio in X-ray band. High harness implies a strong jet and extended hot inner flow, while low hardness implies the dominance of the cold disk. The evolution is schematically illustrated in Fig.~\ref{fig:turtle} (for an actual plot for the outburst of the source GX 339–4 see \citealt{belloni2005}). The source has a jet at the initial stages, then the jet disappears when the cold disk approaches ISCO, and finally the jet reforms. Similar evolution, if taking place in AGN, would take thousands or millions of years \citep{siemiginowska1996}. So having a jet or not might actually be an evolutionary stage.

\section{Broad line region and disk winds in AGN and GBH}

\begin{figure}
\includegraphics[width=\textwidth]{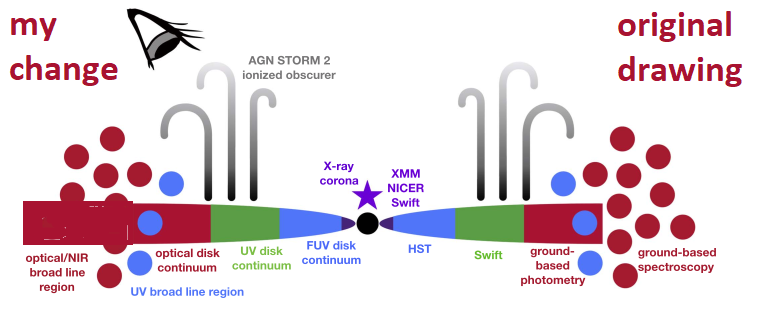}
\caption{The schematic picture of accreting black hole with ouflows; right part follows \citet{Kara2021}, with clouds not overlapping with the disk while left part illustrates the overlap of the clouds and the disk.}
\label{fig:global}
\end{figure}

Accretion onto a black hole is accompanied not only by outflow in the form of a collimated wind but also by uncollimated winds. This is schematically marked in Fig.~\ref{fig:schem2}. This wind is visible directly in absorption (in the form of absorption lines in UV and X-rays). These winds are also seen in emission, as broad emission lines, where the line width is related to the kinematic motion of the emitter (Keplerian rotational velocity plus some outflow component). Such emission lines form the Broad Line Region in AGN, and strong lines are the characteristic features of AGN spectra \citep[see for example a composite quasar spectrum][]{francis1991}, but they are also detected in the optical spectra of some GBH when the data quality is exceptionally good \citep[see for example He I line in MAXI J1820+070 data from VLT/X-shooter][]{rodi2021}. There are arguments that BLR is also directly related to the wind, or a failed wind, from the disk
\citep[e.g.][]{kuraszkiewicz2004,czhr2011}. We illustrate this by changing the original drawing from \citet{Kara2021}, right part of Fig.~\ref{fig:global}, where BLR is outside the disk region, with the setup when BLR and the disk overlap (left part of Fig.~\ref{fig:global}).

\section{Variability of the non-jetted sources}

Accreting black holes are continuously variable in all energy bands. The variability is stochastic and it is related to the way how the excess angular momentum of the material is transported outwards. The key role is played by the magneto-rotational instability \citep{balbus1991}. However, this variability in the cold accretion disk does not yet explain the properties of the power spectra. It is quite likely, that the sandwich model of the coronal flow above the disk can help \citep[e.g.][]{janiuk2007} but more (possibly rather numerical) studies are needed to set this problem.


\begin{figure}
  \centering
  \begin{minipage}{0.48\textwidth}
        \includegraphics[width=\textwidth]{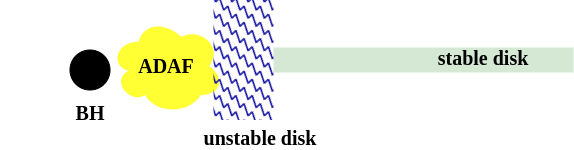}
    \caption{Schematic illustration of the possible mechanism of Changing-Look AGN proposed by \citet{sniegowska2020}.}
    \label{fig:marzena}
  \end{minipage}
  \quad
  \begin{minipage}{0.48\textwidth}
        \includegraphics[width=\textwidth]{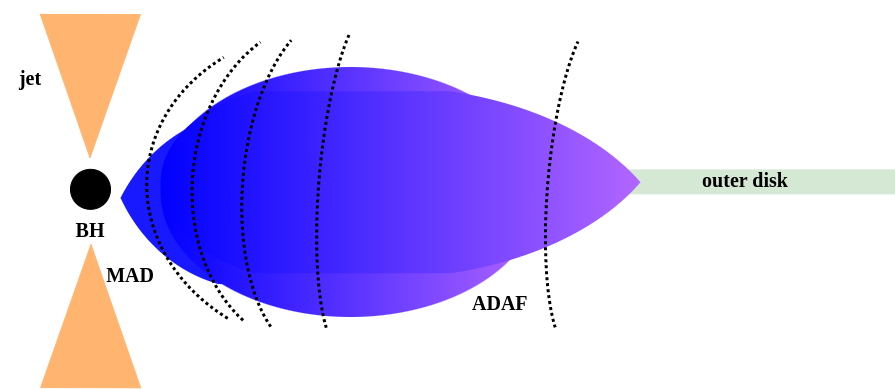}
    \caption{The schematic illustration of the MAD state in the GBH MAXI J1820+070 discussed in \citet{You2023}}
    \label{fig:bei}
  \end{minipage}
\end{figure}

\section{Unsolved problems and complementarity of AGN and GRB studies}

Unsolved problems appear best when we face phenomena not find good explanations, and they well point the holes in the picture presented above. In the case of AGN, this is the phenomenon of the Changing-Look AGN, when in a timescale of months/years an observational appearance of the source changes, going far beyond the usual stochastic variability. Such changes were eventually expected to take centuries or more. We now know about 200 such sources, and some of them went through several such transitions \citep[see e.g.][for a recent study]{panda2022}. In \citet{sniegowska2020} we proposed that this rapid change in the intrinsic flux is caused by a narrow radiatively unstable ring between the cold disk and the hot inner ADAF (see Fig.~\ref{fig:marzena}). There are also other models, and the problem is far from being solved, as the description of the cold disk/ADAF transition is not satisfactory apart from years of attempts. Even more of a problem are Quasi-Periodic Eruptions (QPE; \citealt{miniutti2019}) with the outburst timescale of the order of hours despite the black hole mass $\sim 10^6$ solar mass. Six sources of this class have been discovered so far. Such a phenomenon would last $\sim 10$ seconds in GBH but no analogous outbursts were found so far.

The hot inner flow description also poses some problems. The key issue is the role of the large-scale magnetic field. This field, dragged inward and accumulating close to the black hole horizon can lead to the formation of the magnetically arrested disk (MAD) state. Such a state was predicted theoretically (\citealt{narayan2003}, but see also \citealt{BK1974} for the original idea). In the recent paper \citep{You2023} we directly see not only the evidence of this phenomenon in
the measured time delays, but also how the MAD is formed, in the outburst of a GBH MAXI J1820+070 (see Fig.~\ref{fig:bei} 
for simple illustration.)

There are thus still numerous problems to solve, and that can be done by advanced GR-R-MHD simulations (magnetohydrodynamical simulations which include General Relativity effects, and radiative transfer) and, complementary, by developing simple but still more physical models that can be conveniently fit to the data. Complementary studies of GBH and AGN are very important since in AGN we better resolve the short timescales when they are normalized to the light crossing time of the black hole horizon. In GBH we have, on the other hand, a unique insight into longer timescale evolution. Also, from the spectral point of view, in AGN the cold disk is relatively better studied, apart from the innermost part which is frequently hidden in the unobserved EUV band. In GRH we see the cold disk in highly accreting sources we see the disk, but only the innermost part, and it is frequently strongly affected by the Comptonization in the hot plasma. Combining the efforts we can more easily gain a better understanding of the black hole accretion process.

\acknowledgements{We thank Krzysztof Leszczy{\'n}ski and Micha{\l } Bejger
  for creating and developing this \texttt{ptapap} class. BC acknowledges the support from the European Research Council (ERC) under the European Union’s Horizon 2020 research and innovation program (grant agreement No. [951549]). BY acknowledges the support from the National Science Foundation of China (NSFC) grant 12322307.
  This work was partially supported by the Polish Astronomical Society and the Nicolaus
  Copernicus Astronomical Center, Warsaw, Poland.}

\bibliographystyle{ptapap}
\bibliography{pta4authors}

\end{document}